\documentstyle[12pt]{article}
\begin{document}
\thispagestyle{empty}
\null\vskip 2.cm
\centerline{\Large\bf THE DECAY CONSTANTS }
\vskip0.5cm
\centerline{\Large\bf OF PSEUDOSCALAR MESONS}
\vskip0.5cm
\centerline{\Large\bf IN A RELATIVISTIC QUARK MODEL}
\vskip 1.5cm
\centerline{{\large L.Micu}
\footnote{E-mail address:~~MICUL@THEOR1.IFA.RO}}
\centerline{Department of Theoretical Physics}
\centerline{Institute of Physics and Nuclear Engineering}
\centerline{Institute of Atomic 
Physics P.O.Box MG-6, 76900 Romania}
\vskip 2cm
\begin{abstract}
The decay constants of pseudoscalar mesons
are calculated in a relativistic quark model which assumes
that mesons are made of a valence quark antiquark pair and 
of an effective vacuum like component. The results are given
in terms of quark masses and of some free parameters entering the
expression of the internal wave functions of the mesons. By using
the pion and kaon decay constants
$F_{\pi^+}=130.7~MeV,~F_{K^+}=159.8~MeV$ to fix the parameters of
the model one gets 
$60~MeV\leq F_{D^+}\leq 185~MeV,~95~MeV\leq  
F_{D_s}\leq230~MeV,~80~MeV\leq
F_{B^+}\leq205~MeV$ for the light quark masses
$m_u=5.1~MeV,~m_d=9.3~MeV,~m_s=175~MeV$ and the heavy quark
masses in the range:
$1.~GeV\leq m_c\leq1.6~GeV,~4.1~GeV\leq m_b\leq4.5~GeV$. In the
case of light neutral mesons one obtains with the same set of
parameters $F_{\pi^0}\approx 138~MeV,~F_\eta\approx~130~MeV,F_{\eta'}
\approx~78~MeV$.
The values are in agreement with the experimental data and
other theoretical results.

\end{abstract}
\newpage
\null\vskip 2cm
\section{ Introduction}

The decay constants of pseudoscalar mesons have been treated by
current algebra and PCAC like simple scale parameters relating
the meson fields with the coresponding axial currents. In 
quark models they are expressed by means of the quark-antiquark
annihilation amplitude [1], but, although simple in
principle, the calculation of the decay constants and, in
general, of the electro-weak form factors, is a difficult task
due to the binding effects which escape a relativistic treatment.

A solution is to bypass the binding problem and
work with free quarks. This is the way followed by QCD sum
rules [2], which rely on the assumption of quark-hadron duality
and relate the hadronic matrix elements with some quark and gluon
transition amplitudes which can be evaluated within the
perturbative QCD scheme. This is a fruitful method which
produced most of the recent theoretical results.

Another solution is to find a suitable description of the quark
annihilation in the bound states. The current assumption here is
that the decay constants are proportional with the internal
wave function of the meson at zero distance between the quarks
[1]. Using various forms for the binding potential [3] one gets
that $F$ behaves like $\psi(0)~M^{-1/2}$, where $M$ is the meson 
mass. Potential models work in the case of heavy mesons only,
but they are not Lorentz covariant and are improper in
many cases involving heavy mesons, like, for instance, in 
heavy to light semileptonic decays, where the movement
of the recoiling meson cannot be neglected. For this case and
many others where the hadron movement is important one needs a
relativistic treatment of the binding. Unfortunately, the best
relativistic theory we have at hand, the field theory in the 
perturbative approach, is unable to give an easy answer to the 
binding problem. In our opinion its failure in describing the
bound states is due to the lack of a relativistic equivalent
of the binding energy. We thus suggest to renounce representing 
the binding by a series of some quantum exchanges, since binding
is not a perturbative effect and look instead for a relativistic
generalization of the binding energy which can be included in a
suitable relativistic wave function for the free particles
forming the composite system. In this way we hope to combine the
valuable features of the potential models, which are suitable for
describing the binding, but are  improper for introducing the
boost of the bound state, with those of the relativistic models
which can boost the free states, but cannot describe the binding.

The model we propose has been recently applied to the weak
radiative decays of pseudoscalar mesons [4] and to the decay
$Z^0\to\pi^0\gamma$ [5]. In this paper we intend to exploit
further the properties of the model and to perform some 
predictions on the values of the decay constants of heavy mesons.

The specific assumption of the model is that mesons are made of
a valence $q\bar q$ pair bound together by some collective
oscillation modes of the quark gluonic field. The last ones
are described by an effective vacuum-like component $\Phi$
carying its own 4-momentum which is not subject of a mass-shell
constraint since $\Phi$ is far from being elementary.

As it will be further shown, it is the presence of the field
$\Phi$ which allows to keep the mesons and quarks on their mass
shell while ensuring the energy-momentum conservation together
with a continuous distribution of the quark relative momenta
inside the mesons. One solves in this way the problem mentioned
by Isgur, Scora, Grinstein and Wise [6] who noticed that a
"mock meson" made of almost free quarks with a continuous
distribution of the relative momentum has a false mass width
reflecting the fact that the sum of the free quark momenta does
not belong to a certain representation of the Lorentz group. 

Turning now to the potential models for the meson as a bound
state, we recall that the distribution of the relative momentum
is given by the Fourier transform of the internal wave function.
The existence of a continuous distribution ensuring the $L^2$
integrability of the wave function appears to be a direct
consequence of the binding potential. 

In our model it is the effective field $\Phi$ which adjusts the
continuous distribution of the relative momentum to the free
particle behaviour required by relativistic covariance. This
argument gives a substantial support to the assumption that
$\Phi(Q)$ represents the excitations responsible  for the
confinement and allows us to consider $Q_\mu$  the
relativistic generalization of the potential energy. In fact,
$Q_0$ only is the analogue of the potential energy of the $q\bar q$ 
system at rest. The spatial components {\bf Q} introduced
for relativistic consistency could be rather related with the
fluctuations of the confining oscillation modes of the quark
gluonic field, not with the intensity of the binding forces.

An important ingredient of the model is the internal function of
the compound system representing the hadron.
In the lack of a dynamical equation for it, we shall use some
trial functions allowing to ensure the integrability of the
matrix elements of interest.  

Finally we wish to stress once again that a real relativistic 
treatment of a system made of independent comstituents requires
the use of the momentum space. If the states were defined
in the configuration space, it would be necessary to introduce
an independent time coordinate for each component, which is
nonsense.

In the next section we discuss shortly the dynamical assumptions
of the model and give the expressions of the decay constants as
functions of the quark and meson masses.
 
The numerical results obtained with exponential and gaussian
internal functions are given in the third section. The fit of
the pion and kaon decay constants with the experimental values
is used to fix the parameters of the model. 

We analyse the results in the fourth section and draw some
general conclusions concerning the reliability of the model.
\vskip 1cm
\section{ Calculation of the decay constants}

The fundamental dynamical assumption of the model is that the 
interaction inside the quark system representing a hadron 
can be treated independently from the external interaction. The
first one is a mean field effect and is taken into account 
by means of the internal wave function, while
the external interaction is the effect of some specific quantum 
fluctuations. We recall that this is also the main assumption
underlying the Furry representation in field theory [7].

The form we proposed for the meson state is [4]:
$$\left.\vert M_i(P)\right\rangle=~{i\over(2\pi)^3}~
\int{d^3p\over e/m}
{d^3q\over \epsilon/m'}~d^4Q~\varphi(p,q;Q)~\bar u(p)\Gamma_M
v(q)~\chi^\dagger\lambda_i\psi~\delta^{(4)}(p+q+Q-P)~$$
$$\left.\times\Phi^\dagger(Q)a^\dagger(p)b^\dagger(q)
\vert 0\right\rangle\eqno(1)$$
where $a^\dagger, b^\dagger$ are the creation operators of the
valence $q\bar q$ 
pair; $u,v$ are Dirac spinors and $\Gamma_M$ is a Dirac matrix 
ensuring the relativistic coupling of the quark spins. The quarks 
are supposed to be free; their creation and annihilation operators
satisfy canonical commutation relations and commute with
$\Phi^\dagger(Q)$, which describes the creation of a
nonelementary excitation carrying the momentum $Q_\mu$. The mass
spectrum of the nonelementary excitations denoted by
$\Phi\dagger$ and the internal distribution of momenta are
described by the Lorentz invariant function $\varphi (p,q;Q)$. A
natural assumption is that $\varphi$ is a time independent,
equilibrium distribution since the hadrons are long living.
This means that as long as a quark system like that
described by (1) is the single one in the external state and as long
as it does not emit and absorb any electroweak quanta, the
distribution of momenta is given by $\varphi(p,q;Q)$ and does
not change. A straightforward consequence is that any time
translation operator $U_s(t,t')$ describing the evolution of a
quark system under the action of strong forces only can be
replaced by unity when acting on a state like (1). This fact
will allow us to perform some simplifications in the calculation
of the matrix elements of interest.

For a better understanding of the present model, a comparison
of the expression (1) with the "mock meson" in Ref.[6] is most
useful. A first remark is that the continuous distribution of
the relative momentum in a meson made of a quark and of an
antiquark only, introduced by hand more than 20 years 
ago [6,8], follows naturally in our model from the existence 
of a third component, the field $\Phi$, which contributes to the 
meson momentum. A second remark is that, unlike the "mock meson", 
the expression (1) can be safely boosted due to the function
$\delta^4(p+q+Q-P)$ which guarantees that the sum of the internal
momenta belongs to the representation of the Lorentz group
having the meson mass as invariant.

As concerns the concrete form of the internal function $\varphi
(p,q;Q)$ some more comments are necesary. It must be said that
we have no {\it a priori} arguments for a particular form. We 
expect, however, for $\varphi$ to be such as to ensure the convergence
of the integrals over the internal momenta in the expressions of the
physical amplitudes.

Related to this fact, we remark that if the fourth component of the
momentum carried by the effective field $\Phi$ was positive in the
rest frame of the meson, the function $\delta^{(3)}({\bf p}+{\bf q}
+{\bf Q})~\delta(e+\epsilon+Q_0-M)$ would provide some natural upper 
bound for the quark energies. The integrals in the expression (1)
of a meson state would then extend over a finite range and their
convergence would be easy to ensure.

This is however not the case since, as shown by the electromagnetic
form factors, there is no upper bound for the quark energy in a
hadron. One must then allow for negative values of $Q_0$ and introduce
some definite cut-off functions to ensure the convergence of the
integrals. 

The trial functions we shall use in the next as internal functions 
cut-off the large values of $Q_0$ and {\bf Q} only, but, due to the 
presence of the function $\delta^4(p+q+Q-P)$ we expect for
them to provide the necessary cut-off for the quark momenta too.
A meson at rest is supposed to have an internal function of the
following kind: 
$$\varphi(p,q;Q)~=~D_M\sigma(Q_0,{\bf Q})\eqno(2)$$
$$\sigma(Q_0,{\bf Q})~=~\exp\left[{Q_0\over \alpha}-{\vert{\bf
Q}\vert\over\beta}\right]\theta(Q^2)\theta(-Q_0)~\eqno(2a)$$
$$\sigma(Q_0,{\bf Q})~=~\exp\left[{Q_0\over \alpha}-{{\bf
Q}^2\over \beta^2}\right]\theta(Q^2)\theta(-Q_0)\eqno(2b)$$
$$\sigma(Q_0,{\bf Q})~=~\exp\left[-{ Q_0^2\over \alpha^2}-
{{\bf Q}^2\over\beta^2}\right]\theta(Q^2)\theta(-Q_0)\eqno(2c)$$
where $M$ is the meson mass, $\alpha,~\beta$ are the free
parameters of the model ensuring the desired convergence of the 
integrals. Lorentz covariance of the internal function becomes 
obvious if one writes $Q_0$, and $\vert {\bf Q}\vert$ as: 
$Q_0=(P\cdot Q)/M, \vert {\bf
Q}\vert=\sqrt{(P\cdot Q)^2/M^2-Q^2}$, where $P$ is the meson
momentum. The functions $\theta$ in eqs. (2a), (2b), (2c) expresss the
fact that $Q$ is time like and $Q_0$ negative, in agreement with
our assumption  that $Q_\mu$ is the relativistic generalization
of the potential energy in the bound system.

Before proceeding to the evaluation of the decay constants, we have 
to do an explicit statement on the vacuum expectation value 
of the effective field. As mentioned above, the momentum carried
by $\Phi^\dagger$ is not subject of a mass-shell constraint, since
it represents the creation of a collective excitation, not of an
elementary one. Accordingly, we assume that: 
$$\Phi(Q_1)~\Phi(Q_2)\cdot\cdot\cdot\Phi^+(Q_n)=
\Phi(Q_1+Q_2\cdot\cdot\cdot-Q_n)$$
and that the vacuum expectation value of the effective field
$\Phi(Q)$ is:
$$\left\langle 0\vert\Phi(Q)\vert0\right\rangle~=~\mu^4
\int d^4~X~\exp(-iQ\cdot
X)~=~(2\pi)^4~\mu^4~\delta^{(4)}(Q).\eqno(3)$$ 
We emphasize that the appearance of the function $\delta^{(4)}$ in
(3) is essential for ensuring the overall energy-momentum
conservation and for preserving the Lorentz
covariance of the model. 

The constant $\mu^{4}$ in eq. (3), introduced for 
dimensional reasons, is related with the volume of a large
4-dimensional box of interest for our problem by
$\mu^4=(VT)^{-1}$. A short comment on the box size will
be given in the last section.

Then, using the relations (1) and (3) we get for the norm of a
single meson state the following expression:
$$\left\langle~M_i(P')~\vert M_j(P)~\right\rangle~=~(2\pi)^3~
\delta_{ij}~\delta^{(3)}(P-P') 
\delta(E-E')~(2\pi)^4~\mu^4$$
$$\times\int {d^3~p\over e/m}~{d^3~q\over \epsilon/m'}~d^4~Q
~\delta^{(4)}(p+q+Q-P)\varphi(p,q;Q)^2~Tr\left({\hat p+m
\over 2m}~\Gamma_M~{\hat q-m'\over2m'}~\Gamma_{M'}\right).\eqno(4)$$
The function $\delta^{(3)}(P-P')~\delta(E-E')$ in eq. (4) 
originating from $\delta^{(4)}(p+q+Q-P)$ in the definition of a single 
meson state
can also be written as ${E\over M}~\delta^{(3)}(P-P')\delta(M-M')$.
It reflects the existence of a continuous mass spectrum for
the complex system representing the meson and cannot be modified
without renouncing the real Lorentz invariance of the model.
Its appearance in the expression of the norm forces us to 
treat the physical meson as a mixed state, defined with the aid of 
the diagonal density matrix $\rho(M)$ which satisfies the
normalization condition
$$\int\rho (M)~dM~=~M_0.\eqno(5)$$
where $M_0$ is the central value of the mass distribution.
 Consequently, the density of states in the phase space must be
modified by replacing 
${1\over(2\pi)^3}~{d^3~P\over 2E}$ with ${1\over (2\pi)^3}~{d^3 P
\over 2E}~\rho (M,M_0)~d{M\over M_0}$. If the dependence of
the matrix elements on the meson mass $M$ is rather smooth and 
the width of the mass distribution function is small, one can replace
$M$ by $M_0$ in the expressions of the matrix elements and perform 
the integral over meson masses in the new expression of the density 
of states by using the normalization condition (5).The
calculation can then proceed like in the old case.

The matrix element of interest for the leptonic decay of a
meson, written in the lowest order of
perturbation with respect to the weak interaction is:
$$\left\langle0\left\vert~U_s(+\infty,0)~A_\mu(0)~U_s(0,
-\infty~\right\vert M(P)\right\rangle ~=~i~F_M~P_\mu,\eqno(6)$$
where the operator $U_s(t,t')$ describes the evolution of a system
under the action of strong interaction among the constituents,
and $A_\mu$ is the free-field weak current of interest in the
process. It is important to notice that in soft processes, like,
for instance the present one, the perturbative expansion of
$U_s$ is improper. For this reason we shall not consider the
virtual states generated by the evolution operator in the
perturbative approach, but merely look at the real modifications
which could appear in the distribution of flavours and momenta
during the time translation. In the above case no such changes
could appear, since the real vacuum and the single meson state
are stable states whose content does not change under the action
of strong interaction and consequently both time translation
operators in eq. (6) can be replaced by unity.
 
By using the relation (3), the canonical anticommutation relations 
of the fermionic operators and integrating over the internal momenta,
we obtain from the matrix element (6) the following expession 
for the decay constants:
$$F_M~=~(2\pi)^4\mu^4~D_M~2\pi\sqrt{3}~{p~(m+m')\over M}~\left[1-
{(m-m')^2\over M^2}\right]\eqno(7)$$
where $p={M\over2}~\sqrt{[1-{(m+m')^2\over M^2}]~
[1-{(m-m')^2\over M^2}]}$ and the factor $\sqrt{3}$ comes from the 
colours.

It is worthwhile noticing that the leptonic decay constant in
eq. (7) is proportional with the internal wave 
function at $Q_\mu=0$, which means the absence of any other
excitations beside the valence quarks. Expressing this result in
more general terms, one may say that the leptonic decay
constants are proportional with the value of the internal
function at vanishing contribution from the binding effects.
This is in remarkable agreement with the old assumption that
$F_P$ is proportional with the internal wave function at vanishing
distance between the quarks [1], since, according to the asymptotic
freedom, this is the point in the
configuration space where the confining forces vanish. It is a
strong argument for considering the present model as a real 
relativistic generalization of the potential models.

By integrating now over the quark momenta in
eq.(4) and introducing the expression of the normalization
constant $D_M$ as given by (7), we get the expression of the
decay constants in terms of the model parameters $\alpha,~\beta,
~\mu$ and of the quark masses:
$$F_M~=~(2\pi)^2\mu^2~(12\pi^3)^{1/2}~M(m+m')\left(1-{(m+m')^2\over
M^2}\right)^{1/2} 
\left(1-{(m-m')^2\over M^2}\right)^{3/2}$$
$$\times\left\{\int_{{\bf Q}^2\leq Q_0^2}
dQ_0~{\bf Q}^2~d\vert{\bf Q}\vert\sigma^2(Q_0,{\bf
Q})~\left[{(M-Q_0)^2- 
{\bf Q}^2-(m-m')^2\over M^2~[(M-Q_0)^2-{\bf Q}^2]}\right]\right.$$
$$\times\left.\sqrt{[(M-Q_0)^2-{\bf Q}^2]^2-2
[(M-Q_0)^2-{\bf Q}^2] (m^2+m'^2)+(m^2-m'^2)^2}\right\}^{-1/2}.\eqno(8)$$ 

Similar expressions can be written for the decay constants of
neutral mesons. Defining them like in Ref. [8, p.1444], one has:
$$F_{M^0}~=~(2\pi)^2\mu^2~(24\pi^3)^{1/2}~\sum_{i=u,d,s}~\kappa_i^2~
m_i~\left(1-{4m_i^2\over M^2}\right)^{1/2}$$
$$\times\left\{\sum_{i=u,d,s}~\kappa_i^2~\int~dQ_0~\vert{\bf Q}
\vert d\vert{\bf Q}\vert~\sigma^2(Q_0,{\bf Q})\right.$$
$$\left.\times\sqrt{\left[\left(1+{Q_0\over
M}\right)^2 -{{\bf Q}^2\over M^2}\right]~\left[\left(1+{Q_0\over
M}\right)^2 -{{\bf Q}^2\over M^2}-{4m_i^2\over M^2}\right]}
\right\}^{-1/2} \eqno(9)$$
where $m_i$ are the quark masses, $\kappa_i=a(\lambda_3)_{ii}+
b(\lambda_8)_{ii}+c(\lambda_0)_{ii}$ with $\lambda_j$ the Gell
Mann matrices [8, p.1288], $a=1;~b=c=0$ for $\pi^0$,
$a=0;~b=cos\theta_P;~c=-sin\theta_P$ for $\eta$,
$a=0;~b=sin\theta_P;~c=cos\theta_P$ for $\eta'$ and
$\theta_P=-10^0$ or $\theta_P=-23^0$ [8, p.1320].
\vskip 0.3cm
\section{Numerical results} 

Before proceeding to the numerical calculations we have to
analyse the relation of the model parameters
$\alpha,~\beta,~\mu$, with the general features of
the bound $q\bar q$ system.

First of all we remind that $Q_0$ is the analogue of the
potential energy in the nonrelativistic models. In the present
approach it is the energy of the oscillation modes of the
quark-gluonic field confining the valence quarks inside the
meson. Its cut-off parameter, $\alpha$, must be choosen in
such a way as to ensure a relative stability of $F_P$ with the
increase of $M_P$. (See eq. (8)). Our tests with 
$\alpha=\rho\sqrt{m~m'~M/(m+m')}$,  
$\alpha=\rho\sqrt{\sqrt{m~m'}~M}$, $\alpha=\rho~(m+m')$ and
$\alpha=\rho~M$ where $M$ is the meson mass and $\rho$ a
universal parameter, proved that the last choice is the best. All the
others either lead to very small values for the decay constants
of the heavy mesons or do not allow to fit pion and kaon decay
constants with the same set of parameters, as required from the
begining.
 
The same stability
argument forces us to introduce an additional cut-off for
$\vert{\bf Q}\vert$, the momentum carried by the effective
component, since the simple requirement for $Q^2$ to
be positive would lead to a too strong increase of $F_P$ with
the meson mass. We recall that $\vert{\bf Q}\vert$ has been
introduced for relativistic consistency; we did not relate it
with the potential energy, but rather with some 
fluctuations in the momentum 
carried by the collective excitations denoted by $\Phi$. The
parameter $\beta$ in eqs.(2a,b,c) is hence a measure of the
fluctuation amplitude and we shall assume that it does not depend
on the quark or meson masses because the vacuum-like excitations
are not sensitive to the flavours. However, we expect for
$\beta$ to be smaller than the cut-off parameter of $Q_0$ in the
case of heavy mesons, because the fluctuation effect must be
negligible in their cases.

The parameter $(2\pi)^4\mu^4$ is assumed to be an universal
constant, related with the volume of the 4-dimensional box
relevant for the process. Its independence on the masses will be
used to fix the parameters $\alpha$ and $\beta$ of the model.
Our procedure is to introduce the well known values of the pion
and kaon decay constants $F_\pi=130.7~MeV,~F_K=159.8~MeV$ [9] in
the equation (8) and search for the values of the parameters
$\alpha$ and $\beta$ yielding the same value for $\mu$.
The calculations have been done using the values of the 
light quark masses $m_u=5.1~MeV,~m_d=9.3~MeV,~m_s=175~MeV$
resulting from the chiral perturbation theory [10] and the heavy
quark masses in the range quoted by Particle Data [9].
The results obtained using the trial functions (2a), (2b), (2c) are
listed in the following tables. The quark masses and the decay
constants are given in MeV.
\vskip0.5cm
\noindent
TABLE I. Decay constants of heavy mesons. The indices (a),
(b), (c) correspond to the trial functions (2a), (2b), (2c). 
\vskip0.5cm
\begin{tabular}{c|c c r r c r}
$\alpha_{(i)},~\beta_{(i)}$&\multicolumn{2}{c}{$D(c\bar d)(1869)$}&
\multicolumn{2}{c}{$D_s(c\bar s)(1969)$}&
\multicolumn{2}{c}{$B(u\bar b)(5279)$}\\
$(2\pi)^4\mu^4_{(i)}$&$m_c$&$F_D$&$m_c$&$F_{D_s}$&$m_b$&$F_B$\\
\hline
\hline
$\alpha_{(a)}=0.075~M$&1000.&126.&1000.&152.&4100.&107.\\
$\beta_{(a)}=0.096~MeV$&1300.&109.&1300.&141.&4300.&92.\\
$(2\pi)^4\mu^4_{(a)}=310.~MeV^4$&1600.&56.&1600.&91.&4500.&64.\\
\hline
$\alpha_{(a)}=0.05~M$&1000.&125.&1000.&151.&4100.&111.\\
$\beta_{(a)}=0.065$&1300.&111.&1300.&142.&4300.&95.\\
$(2\pi)^4\mu^4_{(a)}=57.~MeV^4$&1600.&60.&1600.&95.&4500.&77.\\
\hline
$\alpha_{(a)}=0.025~M$&1000.&121.&1000.&144.&4100.&98.\\
$\beta_{(a)}=0.032$&1300.&111.&1300.&139.&4300.&87.\\
$(2\pi)^4\mu^4_{(a)}=3.3~MeV^4$&1600.&65.&1600.&98.&4500.&72.\\
\hline
\hline
$\alpha_{(b)}=0.075~M$&1000.&155.&1000.&188.&4100.&148.\\
$\beta_{(b)}=0.082~MeV$&1300.&137.&1300.&177.&4300.&126.\\
$(2\pi)^4\mu^4_{(a)}=397.~MeV^4$&1600.&73.&1600.&118.&4500.&102.\\
\hline
$\alpha_{(b)}=0.05~M$&1000.&156.&1000.&190.&4100.&153.\\
$\beta_{(b)}=0.054~MeV$&1300.&140.&1300.&180.&4300.&132.\\
$(2\pi)^4\mu^4_{(a)}=71.7~MeV^4$&1600.&79.&1600.&124.&4500.&108.\\
\hline
$\alpha_{(b)}=0.025~M$&1000.&155.&1000.&188.&4100.&156.\\
$\beta_{(b)}=0.027~MeV$&1300.&143.&1300.&181.&4300.&136.\\
$(2\pi)^4\mu^4_{(a)}=4.1~MeV^4$&1600.&85.&1600.&129.&4500.&114.\\
\hline
\hline
$\alpha_{(c)}=0.075~M$&1000.&185.&1000.&227.&4100.&203.\\
$\beta_{(c)}=0.04~MeV$&1300.&168.&1300.&216.&4300.&177.\\
$(2\pi)^4\mu^4_{(c)}=117.~MeV^4$&1600.&95.&1600.&149.&4500.&144.\\
\hline
$\alpha_{(c)}=0.05~M$&1000.&183.&1000.&223.&4100.&204.\\
$\beta_{(c)}=0.027~MeV$&1300.&168.&1300.&214.&4300.&178.\\
$(2\pi)^4\mu^4_{(c)}=22.3~MeV^4$&1600.&99.&1600.&152.&4500.&148.\\
\hline
$\alpha_{(c)}=0.025~M$&1000.&176.&1000.&213.&4100.&198.\\
$\beta_{(c)}=0.014~MeV$&1300.&163.&1300.&207.&4300.&175.\\
$(2\pi)^4\mu^4_{(c)}=1.3~MeV^4$&1600.&99.&1600.&151.&4500.&147.\\
\hline
\end{tabular}
\vskip0.5cm
Using the equation (9) and the same sets of parameters as above
we calculated also the decay constants of the lightest
pseudoscalar mesons. The results are quoted in the next table.
\vskip0.5cm
\noindent
TABLE II. The decay constants of the lightest neutral mesons.
\vskip0.5cm
\begin{tabular}{c|c c c c c}
The trial&$\pi^0(135)$&$\eta(547)$&
$\eta(547)$ &$\eta'(958)$
&$\eta'(958)$\\
function&&$\theta_P=-10^o$ &$\theta_P=-23^o$ &$\theta_P=-10^o$
&$\theta_P=-23^o$\\ 
\hline
\hline
2a&137.-139.&128.-139.&77.-78.&67.-69.&94.-97.\\
2b&$\approx$139.&$\approx$131.&77.-78.&75.-76.&105.-107.\\
2c&135.-139.&129.-132.&76.-78.&77.-81.&108.-114.\\
\hline
\end{tabular}
\vskip0.5cm
\section{Comments and conclusions}

Analysing the numerical results in Table I, one notices that,
for each of the tested internal function, the decay constants do
not change significantly when passing from one set of parameters
$\alpha,~\beta,~\mu$ to another set which fits the values of
$F_\pi$ and $F_K$. Indeed, for a change with 200\% of $\alpha$
and $\beta$, $(4\pi)^4\mu^4$ changes with two orders of
magnitude, while the  theoretical values of the decay constants
change with less than 10\%. 

The decay constants are more
sensitive at the variation of the heavy quark masses and
could be used in principle for a more precise determination of
the last ones. The comparison with the experimental values $F_D\leq
300,~F_{D_s}=232\pm45\pm20\pm48~MeV$, or $F_{D_s}=344\pm37\pm
52\pm42~MeV$ [9, p.1443] and with the values yielded by QCD sum
rules $F_D\approx(1.35\pm0.04\pm0.06)F_\pi$, 
$F_{D_s}\approx(1.55\pm0.10)F_\pi$,
$F_B\approx(1.49\pm0.06\pm0.05)F_\pi$ [11] $F_B=185\pm40~MeV$
[12] shows that 
the agreement is better at the lowest values of the heavy quark
masses. Things look mainly the same for any of the trial
functions, but the best fit of the data seems to be done with
the internal function (2c). Of course, this is just a
qualitative estimate. A more reliable test could be provided by the
fit of the weak or electromagnetic form factors, which are very 
sensitive at the form of the internal function. 

In the case of neutral mesons, we found $(F_{\eta})_{th}$ in the range
$128~MeV-139~MeV$ for $\theta_P=-10^o$, which is in agreement with
$F_\eta=133\pm10~MeV$ quoted in Ref.[9], but
$(F_{\pi^0})_{th}\approx138~MeV$,
slightly larger than $F_{\pi^0}=119\pm4~MeV$ in Ref.[9].
One sees also that the calculated values of $F_{\eta'}$ both for
$\theta_P=10^o$ and $\theta_P=-23^o$ are smaller than
$F_{\eta'}=126\pm7~MeV$, quoted in Ref.[9, p.1444]  
The differences noticed above must not be taken too seriously
because of the large uncertainties entering the values quoted
in Ref.[9]. They come from the 
extrapolation on the meson mass shell when deriving $F_{P^0}$
with the aid of the axial anomaly but, for $\eta,~\eta'$ they
come also from the uncertainties in the mixing angle. 

Resuming, one may say that the present model yields reasonable
values for the decay constants of light {\it and} heavy mesons
using the same set of parameters, which is quite remarkable. 

A last comment concerns the parameter $\mu$.
As it can be seen from Table I, its values resulting from the
fit of the pion and kaon decay constants are in the range
$0.2-0.7~MeV$. Recalling that $\mu^{-4}$ is equal to the
4-dimensional volume $VT$ of a very large box containing the
meson, we get a box size of about $300-1000~fm$,  
quite large in comparison with the meson size which is less than
$1~fm$. The large value found for the box size, as well as the
relative independence of the results in Tables I and II on the
value of $\mu$, if $\mu$ is sufficiently small, are strong
arguments for the consistency of the present model.  
\vskip0.5cm
{\Large\bf References}
\vskip0.5cm
\begin{enumerate}
\item R. van Royen and V.F. Weisskopf Nuovo Cim. {\bf50}, 617 (1967). 
\item For a review of QCD sum rules in the context of heavy
quark effective theory see M. Neubert Phys. Rep. {\bf245}, 259 (1994).
\item E. Eichten, K. Gottfried, T. Kinoshita, T.D. Lane and
T.-M. Yang, Phys. Rev. D{\bf17}, 3090 (1978); C. Quigg and J.L. Rosner,
Phys. Lett. {\bf71B}, 153 (1977); J.L. Richardson, Phys. Lett. {\bf82B},
272 (1979); A. Martin, Phys. Lett. {\bf93B}, 338 (1980); E.J. Eichten 
and C. Quigg, Phys. Rev. D{bf49}, 5845 (1994); S.S. Gershtein, V.V. 
Kiselev, A.K. Likhoded, A.V. Tkabladze, Phys. Rev. D{\bf59}, 3613 (1995).
\item D. Ghilencea and L. Micu Phys. Rev. D{\bf52}, 1577 (1995).
\item L. Micu Phys. Rev. D{\bf53}, 1 (1996).
\item N. Isgur, D. Scora, B.W. Grinstein and M. Wise Phys. Rev.
D{\bf39}, 799 (1989).
\item W.H. Furry, Phys. Rev. {\bf81}, 115 (1951). 
\item A. Le Yaouanc, L. Oliver, O. P\` ene and J.-C. Raynal,
Phys. Rev. D{\bf9}, 2636 (1974); M.J. Ruiz, Phys. Rev. D{\bf12}, 
2922 (1975); N. Isgur, Phys. Rev. D{\bf12}, 3666
(1977).
\item Particle Data Group, Phys. Rev. D{\bf50}, 1288, 1320,
1443 (1994). 
\item H. Leutwyler CERN-TH/96-25; hep-ph/9602255. 
\item E. Bagan, P. Ball, V.M. Braun and H.G. Dosh, Phys. Lett.
B{\bf278}, 457 (1992); M. Neubert, Phys. Rev. D{\bf45}, 2451
(1992). 
\item S. Narison, Acta Physica Polonica B{\bf26}, 687 (1995).
\end{enumerate}
\end{document}